\documentstyle[prl,aps,twocolumn]{revtex}

\begin{document}
\draft
\preprint{ March 10 '98}
\title{Magnetoconductance noise and irreversibilities
in submicron wires of spin-glass n$^+$-Cd$_{1-x}$Mn$_x$Te}
\author{J. Jaroszy\'nski, J. Wr\'obel,
G. Karczewski, T. Wojtowicz, and T. Dietl}
\address{Institute of Physics, Polish Academy of Sciences,
al.~Lotnik\'ow 32/46, PL-02668 Warszawa, Poland}

\maketitle
\begin{abstract}
Signatures of spin-glass freezing such as the appearance of
$1/f$ conductance noise, the recovery of universal conductance
fluctuations, aging, as well as magnetic and thermal
irreversibilities are detected in mesoscopic wires of
Cd$_{1-x}$Mn$_x$Te:I at millikelvin temperatures. Spectral
characteristics of conductance time series are consistent with
the droplet model of short-range spin-glasses.
\end{abstract}
\narrowtext 
Since the seminal suggestion of Altshuler and Spivak
\cite{Alts85} and of Feng {\it et al.} \cite{Feng87} that the
sensitivity of quantum interference of scattered electron waves
to the instantaneous configuration of the localized spins in
mesoscopic systems might serve as an important tool for testing
models of spin glasses \cite{Mydo95}, considerable efforts have
been made to fabricate and study metallic nanostructures doped
with magnetic impurities. In particular, Israeloff {\it et al.}
\cite{Isra89} detected $1/f$ conductance noise in Cu$_{1-x}$Mn$_x$
at temperatures above 5 K.  A detailed analysis of those data by
Weissman and co-workers \cite{Weis92} has demonstrated the
magnetic origin of the noise at and below the freezing
temperature $T_g$ as well as a satisfactory agreement of its
magnitude with the Feng {\it et al.} \cite{Feng87} theory.
Moreover, the results were shown \cite{Weis92} to point to the
hierarchical model of the spin-glass phase \cite{Bind86}, a
conclusion drawn from the character of non-gaussian behavior of
the conductance time series in mesoscopic wires of
Cu$_{0.91}$Mn$_{0.09}$.

After this initial progress,  {\em no} mesoscopic signatures of
spin freezing were found by both Van Haesendonck {\it et al.}
\cite{VanH89} and Benoit {\it et al.} \cite{Beno92}. This was
attributed to a reduction of the strength of the
Ruderman-Kittel-Kasuya-Yosida (RKKY) interactions by disorder
inherent to nanostructured samples and to  a corresponding
increase of the role of the Kondo effect \cite{Neut96}.  It has
been also pointed out that the total number of the localized
spins in the studied structures might be too small for the phase
transformation to show up \cite{Beno92}.  By contrast,
a rather robust spin freezing was observed by de Vegvar
and co-workers \cite{deVe91} in nanostructures of Cu:Mn.  In
particular, an antisymmetric term in the magnetoresistance
tensor, generated by the frozen spins \cite{Hers91}, persisted
above $T_g$ characteristic for the bulk material.  This
surprising observation, together with a visible reduction of the
Kondo resistivity, were taken as indicative of importance of
magnetic inclusions, such as MnO \cite{Weis92b}.

We report here on a study of the resistance noise and magnetic
field-induced fluctuations in submicron wires of a diluted
magnetic semiconductor (DMS) \cite{Diet94}
Cd$_{1-x}$Mn$_{x}$Te:I with electron densities greater than that
corresponding to the metal-insulator transition. In DMS, the
localized spins are coupled by short-range antiferromagnetic
superexchange interactions, which in the studied range of Mn
concentrations, $0.07 \le x \le 0.2$, lead to the spin-glass
transition at $0.3 \lesssim T_g \lesssim 2.2$~K \cite{Nova85},
respectively. Owing to a large difference between the relevant
length scales, the studied wires are mesoscopic from the point
of view of the electronic properties but macroscopic as far as
the range of magnetic interactions is concerned.  This 
feature, together with the absence of the competing Kondo effect
(due to the ferromagnetic character of the $s$-$d$ exchange
interaction \cite{Diet94}), make DMS particularly suitable for
the meaningful examination of the spin-glass phase by the
phenomena of coherent transport.  Our results corroborate the
existence of $1/f$ magnetic noise, detected here down to 30 mK.
At the same time, we observe other signatures of the spin-glass
freezing such as aging, thermal and magnetic irreversibilities
as well as a strong increase in the amplitude of both UCF and
the noise when the temperature {\em and} the magnetic field are
reduced below the freezing line.  We demonstrate that the
present data are consistent with the droplet model \cite{Fish88}
of short-range spin-glasses as well as with qualitative
predictions of Monte Carlo simulations of conductance in
Edwards-Anderson spin glasses \cite{Ciep91}.

According to the theoretical model of mesoscopic phenomena
\cite{Alts85,Feng87,Alts86,Bobk90}, out of eight contributions
to var$(\Delta G)$, seven are diminished by the electron
interaction with the Heisenberg spins virtually independently of
spin dynamics,  while the remaining term varies in tact with the
changes of the local spin configurations.  As a results, the
spins act as phase breakers at time scales shorter than the time
constant of the resistance meter, while they are a source of
equilibrium conductance noise or irreversibilities at longer
epochs.  In particular, the Fourier power spectrum $S(f)$ of the
stationary noise is proportional to the Fourier transform of the
normalized spin-spin autocorrelation function  $C(f) =
2\chi\prime\prime(f,T)/\pi f\chi(T)$ \cite{Feng87,Bobk90}.
Particularly interesting is the case when the noise is
controlled by a small number of fluctuators so that the
conductance time series may exhibit deviations from gaussian
statistics. Spectral characteristics of these deviations, the
so-called second spectrum $S_2(f)$ \cite{Weis92,Rest85,Seid96a},
is white when the fluctuators act independently but assumes a
nonwhite character for either interacting fluctuators
\cite{Weis92} or dynamic redistribution of the current between
its possible paths \cite{Seid96b}. Thus, $S_2(f)$ may provide
information on whether glassy dynamics reflect wandering of spin
configurations between local free energy minima (hierarchical
dynamics) or rather a spontaneous formation and annihilation of
compact droplet excitations \cite{Weis92}.

Our Cd$_{1-x}$Mn$_x$Te:I films with $x=0.07$ and $0.20 \pm
0.005$, having thickness of 0.3 $\mu$m, and electron
concentrations $n$ of about $2\times 10^{18}$ cm$^{-3}$ were
grown by MBE on SI GaAs with 10 {\AA} ZnTe and 3~$\mu$m CdTe
undoped buffer layers.  The magnitudes of $n$ and $x$ were
determined from the room temperature Hall data and the spectral
position of the photoluminescence line, respectively. It has
been confirmed in course of this work that the electrical
activity of iodine impurities \cite{Waag93}, in stark contrast
to indium donors used previously \cite{Jaro95}, varies little
with $x$ for $x \le 0.3$.  Accordingly, the highest values of
$n$, which we obtain by iodine doping is---in the case of
Cd$_{0.8}$Mn$_{0.2}$Te---by four orders of magnitude greater
than that which could be reached with indium doping.  The wires
were fabricated by means of electron-beam lithography, followed
by wet etching in 0.05\% solution of Br$_2$ in ethylene glycol.
As shown in the inset to Fig.~1, they have a mean width of $W =
0.3$~$\mu$m,  and the arrangement of the contacts suitable for
five-probe measurements of conductance noise $G(t)$ by the
a.c.~method \cite{Scol87}. In order to avoid electron heating
\cite{Henn97}, current intensity as low as 100 pA was employed,
which limited the studied range of $1/f$ noise to $f \le 1$ Hz.
The output filters served as an anti-alias device, and they were
set to reject the output voltage components of frequency greater
than $f = 0.3$ and $3$ Hz in the case of measurements {\it vs.}
$H$ and $t$, respectively. A typical field sweep rate was $dH/dt
= 0.5$ kOe/min, and ten 12 bit data points were acquired per
second.  The Kaiser and square windows were employed for the
determination of the first and second Fourier power spectra,
respectively.

Figures 1 and 2 present the conductance $G(H,t)$ in the wires
with $x = 0.07$ and $0.2$, respectively. Similarly to previously
studied paramagnetic Cd$_{0.99}$Mn$_{0.01}$Te:In \cite{Jaro95},
in the present case the electron transport is found to be
significantly affected by the giant and temperature dependent
spin-splitting of the conduction band \cite{Diet94}.  In
particular, the spin-splitting leads to the weak-field positive
magnetoresistance \cite{Sawi86}, visible as a dip near $H=0$ in
Figs.~1(a) and 2(a). At the same time, the redistribution of the
electrons between the spin subbands, and the associated changes
in the length of the interfering waves of the carriers at the
Fermi level, appears to constitute the dominant mechanism of UCF
generation in any magnetic material \cite{Jaro95}. The latter
accounts for a monotonic shift with temperature of the field
values corresponding to given conductance features, an effect
clearly seen in Fig.~1(a).  Below, we shall focus, however, on
those phenomena which are specific to the spin-glass phase.

Starting with the data for n$^+$-Cd$_{0.93}$Mn$_{0.07}$Te wire
(Fig.~1), we note that the UCF amplitude rms($\Delta G$), in low
magnetic fields and at $T > 0.3$~K is weakly temperature
dependent and smaller than that found in similar wires of
n$^+$-Cd$_{1-x}$Mn$_{x}$Te, $x \le 0.01$ \cite{Jaro95}.
However, rms($\Delta G$) is seen to increase abruptly below
0.3~K, temperature corresponding to $T_g$ in the bulk material
with $x=0.07$.  In the same temperature range, a dramatic
increase in the conductance noise is observed.  Our findings
reveal, therefore, a destructive effect of the fluctuating spins
on the UCF, and the appearance of the low frequency noise when
the spin dynamics become slow, corroborating qualitatively
expectations of theoretical models \cite{Feng87,Bobk90} and
numerical simulations \cite{Ciep91}.

As shown in Fig.~2, particularly strong and complex signatures
of the spin-glass freezing are found in the case of
n$^+$-Cd$_{0.8}$Mn$_{0.2}$Te wire.  Such a sensitivity of the
conductance to spin configurations in the whole temperature
range below 1~K stems from a large value of $T_g \approx 2.2$~K
in this material.  Additionally, relatively large chemical and
spin disorder for $x=0.2$ brings the electron liquid close to
the localization boundary and, thus, makes it particularly
sensitive to local variations in the scattering potential. Some
relevant examples of history-dependent effects are depicted in
Fig.~2(a), which shows a series of magneto-fingerprints,
measured in succession as a function of the magnetic field (and
time) after a heat pulse and subsequent cooling from $T > T_g$
to $T \approx 0.02 T_g$. The effect of aging, that is, a gradual
decrease in both the fluctuation amplitude and the differences
between subsequent traces is clearly visible.  Magnetic
irreversibilities persist even after long waiting time. Together
with low frequency equilibrium noise, they make the correlation
coefficient to be as low as 0.7 for the two most lately
registered traces of Fig.~2(a). According to spectral densities
$S(f)$ presented in Figs.~3(a) and 3(b), the noise is white at
$T > T_g$. Below $T_g$, however, $S(f)$ is seen to assume the
form $1/f^{1 +\gamma}$, where at 50 mK and in $H=0$, $\gamma =
0.3$ and $0.5$ for $x=0.07$ and $0.2$, respectively.
 
An important aspect of our data is that they provide information
on how the magnetic field affects the spin-glass dynamics.  It
may appear that the field will reduce fluctuations of the Mn 
spins and, thus, will result in an increase of the UCF
amplitude. Indeed, such an increase has been found by others 
\cite{Beno92} and by us in the paramagnetic phase. 
At $T < T_g$, however, rms$(\Delta G)$ of
both the UCF [Figs.~1(c) and 2(c)] and noise [Fig.~2(b)] is
observed to {\em decrease} when the magnetic field gets
stronger. The former demonstrates an increase of integrated
$S(f)$ in the frequency range $f \ge 0.3$ Hz, while the latter
its decrease for $f \le 3$ Hz. At the same time, according to
results on $S(f)$ at 50 mK presented in Figs.~3(a) and 3(b),
$\gamma$ decreases from 0.3 and 0.5 in $H=0$ to 0.1 and
0.2 in 36 kOe, for $x=0.07$ and $0.2$, respectively. This
means that the maximum of $\chi\prime\prime(f)$ shifts in the
magnetic field to the range of frequencies greater than those
explored in the present experiment.  Thus, our findings lead
consistently to the conclusion that the principal effect of the
magnetic field on spin-glass freezing is to displace the
spectral weight of the magnetic excitations toward higher
frequencies.

As we have already noted, of particularly relevance is the
analysis of non-gaussian effects in noise statistics.  In the
case of the wire with $x = 0.07$, the normalized variance of
$S(f)$ for seventy successively collected data trains, and
average over four octave bands of $f$, is $s_2 = 1.1 \pm 0.2$ at
50 mK. For the adopted normalization, this indicates that the
noise is essentially gaussian, a conclusion corroborated by the
frequency dependence of correlation between the amplitude and
phase components of the Fourier transform of $G^2(t)$, {\it
i.e.}, $s_f^{(2,a)}$ and $s_f^{(2,\phi)}$ \cite{Seid96a}, the
latter depicted in Fig.~3(c).

In contrast to the data for $x = 0.07$, a stronger freezing
specific to $x= 0.2$, together with the proximity of the
metal-to-insulator transition, result in significant
non-gaussian effects. In particular, the value $s_2 = 16 \pm 2$
at 100 mK [Fig.~2(d)], indicates that only a few fluctuators
controls the noise over the studied frequency range
\cite{Weis92}. It turns out that non-gaussian statistics result
from phase correlation, so that $s_f^{(2,\phi)}$ is plotted in
Fig.~3(c). Of particular significance is its week frequency
dependence, $s_f^{(2,\phi)} \sim f^{-0.3}$.  Actually, despite
an expected contribution of the dynamic current redistribution
in the vicinity of the localization threshold \cite{Seid96b},
the frequency dependence of $s_f^{(2)}$ in Cd$_{0.8}$Mn$_{0.2}$Te
is {\em weaker} than in metallic Cu$_{0.91}$Mn$_{0.09}$ 
\cite{Weis92}, as shown in  Fig.~3(d). This important 
observation implies that the noise in DMS is
dominated by a set of independent fluctuators, presumably,
droplet-like excitations.  The comparison of those two material
systems provides, therefore, the experimental indication that
the droplet model \cite{Fish88} describes slow dynamics more
accurately in the case of the short range superexchange 
interaction than when the coupling between the spins proceeds 
{\it via} the long range RKKY mechanism.

In summary, our results demonstrate that on the scale of the
coherence length of the electron wave function in DMS, the
magnetic subsystem remains unperturbed by the growth or
nanostructuring process.  This has made possible to observe, by
means of quantum phenomena, effects of the temperature and the
magnetic field on equilibrium and non-equilibrium glassy
dynamics, providing the important verification of the
theoretical predictions \cite{Alts85,Feng87,Bobk90} and of the
numerical simulations \cite{Ciep91}.  In view of our findings,
the absence of both the conductance noise
\cite{VanH89,Beno92,deVe91} and the field-induced
irreversibilities \cite{deVe91} in mesoscopic metallic
spin-glasses might indeed result from finite-size and clustering
effects.  Statistical analysis of the conductance time series
indicates that in contrast to diluted magnetic metals
\cite{Weis92}, DMS constitute the material system, to which the
description of spin-glass properties in terms of droplet
excitations \cite{Fish88} may directly apply. Our results
demonstrate also that the magnetic field shifts gradually the
spectral weight of these excitations toward higher frequencies.

We thank J. Chroboczek and J. Kossut for discussions, and KBN
for support under Grant No. 2-P03B-6411.

\begin{figure}
\caption[]{Conductance $G$ as a function of the magnetic field
$H$ (a) and time at $H = 0$ (b) in the wire of
$n^+$-Cd$_{0.93}$Mn$_{0.07}$Te at selected temperatures down to
30 mK (uppermost traces). Note that $G$ decreases in the upward
direction.  Inset to (b) shows atomic force micrograph of the
sample. Temperature dependencies of the root mean square
conductance fluctuations for $x = 0$, $0.01$ \cite{Jaro95}, and
$0.07$ (empty symbols) and noise for $x = 0.07$ (full symbols)
are shown in (c); solid lines are guide for the eye.  Arrow
marks the bulk value of the spin-glass freezing temperature for
$x = 0.07$.}
\end{figure}

\begin{figure}
\caption[]{Conductance $G$ as a function of the magnetic field
(a) and time (b) in the wire of $n^+$-Cd$_{0.8}$Mn$_{0.2}$Te at
50 mK.  The traces in (a) were taken in succession starting from
the uppermost, and are shifted by $1e^2/h$ for clarity. Arrows
indicate directions of the sweep.  Note that $G$ decreases in
the upward direction. Temperature dependencies of the root mean
square conductance fluctuations (empty symbols) and noise 
(full symbols) are shown in (c), while fractional variance of 
the noise power spectrum in (d); solid lines are guide for the 
eye.}
\end{figure}

\begin{figure}
\caption[]{Fourier power spectra of noise, $S(f)$ in the wires 
of Cd$_{0.93}$Mn$_{0.07}$Te (a) and Cd$_{0.8}$Mn$_{0.2}$Te (b)
at selected temperatures and magnetic fields.  Normalized second 
noise spectra $s_f^{(2,\phi)}$ taken in the bandwidth from 0.1 to 
0.6 Hz at 50 mK are compared in (c) to expectations for the 
gaussian $1/f$ noise \cite{Seid96a}. The frequency dependence 
of $s_f^{(2)}$ (normalizing to and subtracting the gaussian 
expectation) is shown in (d) for Cd$_{0.8}$Mn$_{0.2}$Te at 
50 mK and Cu$_{0.91}$Mn$_{0.09}$ at 11 K \cite{Weis92}. 
Lines in (a) and (b) show $1/f^{1 + \gamma}$ dependence, 
while in (c) and (d), except for the gaussian background, 
are guide for the eye.}
\end{figure}
\end{document}